\documentclass{elsartm}
\usepackage{graphicx}
\usepackage{amssymb}
\begin{document}

\begin{frontmatter}

\title{Neighborhood properties of complex networks}
\author{Roberto F. S. Andrade, Jos\'e G. V. Miranda}
\address{Instituto de F\'{\i}sica - Universidade Federal da Bahia \\
40.130-240 - Salvador - Brazil}
\author{ and Thierry Petit Lob\~{a}o}
\address{
Instituto de Matem\'{a}tica -  Universidade Federal da Bahia,\\
40210-340 - Salvador - Brazil}

\date{\today}% It is always \today, today,

%\vspace{0.2in}
%\end{center}
\begin{abstract}

A concept of neighborhood in complex networks is addressed based
on the criterion of the minimal number os steps to reach other
vertices. This amounts to, starting from a given network $R_1$,
generating a family of networks $R_\ell, \ell=2,3,...$ such that,
the vertices that are $\ell$ steps apart in the original $R_1$,
are only 1 step apart in $R_\ell$. The higher order networks are
generated using Boolean operations among the adjacency matrices
$M_\ell$ that represent $R_\ell$. The families originated by the
well known linear and the Erd\"os-Renyi networks are found to be
invariant, in the sense that the spectra of $M_\ell$ are the same,
up to finite size effects. A further family originated from small
world network is identified.

\end{abstract}
\end{frontmatter}
%\PACS {89.75.Hc, 89.75.Fb, 89.20.Hh, 02.10.Ox}

%\verb+\pacs{#1}+

Several properties of complex networks have been addressed
recently, much inspired by the identification of their relevance
in the description of the relations among the individual
constituents of many systems, which have their origin from natural
to social sciences, telephone and internet to energy supply, etc.
\cite{Watts,Barabasi,Doro1,Vespignani}. The huge development in
this research area came along with the proposition of many
parameters that might be relevant to the characterization of
properties of networks. The sofar most established and quantifiers
are the distribution of links $n(k)$, clustering coefficient $C$,
mean minimal distance among the nodes $<d>$, diameter $D$, the
assortativity degree \cite{Barabasi02,assortativity}. The
evaluation of these and other indices for actual networks help
characterizing them, putting into some network classes with well
defined properties, such as the small-world \cite{Watts98}, the
scale-free \cite{Barabasi99}, the Erd\"os-Renyi \cite{Renyi} or
random networks, etc. %{In particular, the scale-free character of
%a network expresses how the relative presence of nodes with $k$
%connections decays as $k$ increases\cite{Barabasi99}. This
%indicates, e.g., that if one removes the largest hub in a
%scale-free network, the rest of the network still has the same
%character, the next largest hub assuming the hole played by that
%one just eliminated.}

As the number of nodes directly connected to node $i$ is $k_i$,
$n(k)$ characterizes the immediate neighborhood of the network
nodes. In this work we explore further neighborhood
properties of networks, which are related to the distribution of
the number of second, third, ..., neighbors. For the sake of
simplicity, we assume that the networks we consider herein are
connected, i.e., each node can be reached from any other one. Two
nodes are $\ell-th$ neighbors, or neighbors of order $O(\ell)$,
when the minimal path connecting them has $\ell$ steps. Then, for
a given network $R$, the explicit evaluation of the distributions
of neighbors $O(\ell)$ along the network, $\ell=1,2,3,...$
promptly indicates the structure of minimal paths connecting the
nodes. This classifies uniquely the neighborhood of a vertex, in
the sense that if two vertices are $O(\ell)$ neighbors, they are
not $O(j)$ neighbors for $\ell \neq j$. Also, we consider that any
vertex is $O(0)$ neighbor of itself.

It is expected that the neighborhood properties change with
$\ell$. However, if a meaningful criterion $G$ can be devised,
according to which the $\ell$-neighborhoods of $R$ remain
invariant, it may be important to assign $R$ a $G$ neighborhood
invariant (NI) property. It characterizes an invariance of $R$
with respect to length scale, provided length is measured by
number of steps along the network. Also, it is distinct both from
the scale free distribution of vertex connections, as well as from
geometrical scale invariance included in the construction of the
network as, for instance, in the class of Apollonian networks
(AN).

In our investigation  we identified, for each $\ell$, all
$O(\ell)$ neighbors of $R$ and constructed a family of networks
${\mathbf{R}}$. Each $R_\ell \in \mathbf{R}$ is defined by the
same set of nodes, while links are inserted between all pairs of
vertices that are $O(\ell)$ neighbors in $R\equiv R_1$. Thus, the
family $\mathbf{R}$ characterizes the neighborhood structure of
$R$. The $R_\ell$'s were actually set up by a corresponding family
${\mathbf{M}}$ of adjacency matrices (AM) ${M_\ell}$
\cite{Barabasi02}, achieved by the systematic use of Boolean ($B$)
operations \cite{Boolean} among matrices.

To define a neighborhood invariance criterion, it is necessary to
identify a relevant network property that may be present in all
elements of ${\mathbf{R}}$, e.g., the corresponding values for
$n_\ell(k)$ and $C_\ell$ \cite{Fronczak}. However, these are
global indices that do not provide a sufficiently precise
identification of a network. A much more precise characterization
is based on the eigenvalue spectra of the family
${M_\ell}$\cite{Bolobas,Dods}, even if it is well known that there
may exist topologically distinct networks which share the same
spectrum (co-spectral), and that only the complete set of
eigenvalues and eigenvectors univocally characterizes a network. A
spectrum based invariance criterion condenses (in the sense of
reducing from $N^2$ elements of $M$ to $N$ numbers) a lot of
information about the network, and has been used herein. This can
be further justified by the fact that, if two AM's in ${M_\ell}$
have the same spectrum, they differ only by a basis
transformation.

Before proceeding with details of the guidelines of our actual
work, we recall that the first matrix $M_1$, which describes the
original network $R_1$, has only 0's or 1's as entries for all of
its elements. If $M_1$ is applied to a unitary vector $v_{\ell}$,
with all but one entry $\ell$ set to 0, the resulting vector
expresses which vertices are linked to the the vertex $\ell$. If
we take the usual matrix product of $M_1$ by itself, the non-zero
elements $(M_1^2)_{rs}$ of the resulting matrix $M_1^2$ indicates
how many possible two-step walks along the network, with endpoints
$r$ and $s$ exist. Contrary to what happens with $M_1$, $M_1^2$
has many elements $(M_1^2)_{rs}>1$, indicating multiplicity of
paths starting at $r$ and ending at $s$. In particular, all
elements $(M_1^2)_{rr}$ of the diagonal can have this property,
since they count all two-step walks that start at $r$, visit any
of the vertices $s$ to which $r$ is linked $(M_1)_{rs}=1$, and
turn back to $r$. The same interpretation is valid for all usual
powers of $M_1$.

As the elements of $M_1$ are all 0's or 1's, we can regard
$(M_1)_{rs}$ as $B$  variables, and use the $B$ sum, subtraction
and product operations \cite{Boolean}, respectively $\oplus,
\ominus, \otimes$,
\begin{equation}\label{eq1}
  \begin{array}{llll}
    0 \oplus 0 = 0 & 1 \oplus 0 = 1 & 0 \oplus 1 = 1 & 1 \oplus 1 = 1 \\
    0 \ominus 0 = 0 & 1 \ominus 0 = 1 & 0 \ominus 1 = 0 & 1 \ominus 1 = 0 \\
    0 \otimes 0 = 0 & 1 \otimes 0 = 0 & 0 \otimes 1 = 0 & 1 \otimes 1 = 1
  \end{array}
\end{equation}
to define $B$ operations between matrices of $B$ elements. The $B$
matrix operations are defined by using the usual matrix element
operation rules, replacing the usual sum, subtraction and product
involving matrix elements by the corresponding $B$ operations. To
avoid multiplicity of notation, we will use hereafter the same
symbols $\oplus, \ominus$ and $\otimes$ to indicate matrix $B$
operations.

If we consider $\overline{M_2}=M_1\otimes M_1$ and compare it to
$M_1^2$, we see that the position of all their zero elements
coincides, while if we collapse to 1 all non-zero elements of
$M_1^2$ we obtain $\overline{M_2}$. In fact, $\overline{M_2}$
indicates the possibility of two-step walks, while it deletes the
information on the multiplicity of walks. As the neighborhood
concept does not take path multiplicity into account,  the $B$
operations are well suited to define the matrices $M_i$.

For instance, $M_2$ can be expressed by
\begin{equation}\label{eq2}
M_2=(M_1 \oplus \overline{M_2})-(I\oplus M_1)=(I\oplus M_1)\otimes
M_1 - (I\oplus M_1),
\end{equation}
where $I$ indicates the identity matrix. To see this note that all
forward-backward walks, included together with pairs of distinct
sites linked by two-step walks in $\overline{M_2}$, can not be
present in $M_2$, as any vertex has been defined to be $O(0)$
neighbor of itself. Thus we must subtract $I$ from
$\overline{M_2}$. Also, it is necessary to $B$ sum $M_1$ to
$\overline{M_2}$, and subsequently subtract $M_1$, as
$\overline{M_2}$ may describe two-step walks between two sites
that were already related in the original network $R_1$. Noting
that $M_0\equiv I$, Eq.(\ref{eq2}) can be generalized for
arbitrary value of $\ell$ by:
\begin{equation}\label{eq3}
M_\ell=(\bigoplus_{j=0}^{\ell-1}M_j)\otimes M_1 -
(\bigoplus_{j=0}^{\ell-1}M_j)=M_{\ell-1}\otimes M_1 \ominus
(\bigoplus_{j=0}^{\ell-1}M_j).
\end{equation}

Once a precise procedures to set up all $M_\ell$'s is available,
let us briefly comment on several possibilities opened by the
knowledge of ${\mathbf{M}}$ for the purpose of evaluation of
network indices. 1) If we have a finite network with $\mathbf{N}$
vertices, then there is a large enough $\ell_{max}$ such that
$M_\ell\equiv 0, \forall \ell>\ell_{max}$. Thus, the value for $D$
is found when the first $M_\ell\equiv 0$ is found. Also, when
$\ell$ approaches $D$, the $M_\ell$'s become sparser and, as a
consequence, the number of zero eigenvalues increases largely. 2)
As for each $r,s$ pair, $(m_\ell)_{rs}=1$ for only one
${M_\ell}\in \mathbf{M}$, one can collapse in a single matrix
\begin{equation}\label{eq4}
\widehat{M}=\sum_{j=0}^{\ell_{max}}j M_j
\end{equation}
all information on the neighborhood of any pairs of vertices.
Particularly, all pairs $r,s$ of the $O(\ell)$ neighbors satisfy
$\widehat{M}_{rs}=\ell$. 3) To obtain the average minimal path for
each node $r$, it is sufficient to sum all elements of the $r-th$
row (or column) of $\widehat{M}$ and divide by $\mathbf{N}-1$. The
average minimal path for the network follows immediately. 4) The
evaluation of all $C_\ell$ by means of line to line multiplication
of matrix elements, can be easily computed once the family
${\mathbf{M}}$ has been obtained. 5) The matrix $\widehat{M}$ can
be easily used to visualize the structure of network with the help
of a color or grey code plots.

The numerical evaluation of the eigenvalue spectra
${\lambda_\ell^r,r\in[1,\mathbf{N}]}$ for the family ${M_\ell}$
has been carried out for several networks. For each one of them,
we are particularly interested to understand how the form of the
spectral density $\rho_\ell(\lambda)$ as function of $\lambda$
evolves with $\ell$. We first illustrate the procedure by showing
how the spectra of the standard AN depends on $\ell$. Several
properties of AN's have been discussed recently
\cite{Andrade,Doye}: they are constructed according to a precise
geometrical iteration rule that grant them topological
self-similarity and a scale free distribution of nodes degrees.
The $M_1$ spectra of the AN for $g$ successive generations
converge very quickly to a well defined form. However, for the
finite size networks up to $g=8$, with 3283 vertices, we can not
identify any sort of invariance in the eigenvalues density
distribution $\rho_\ell(\lambda)$. This is better visualized with
the help of the integrated spectra \cite{Miranda}
\begin{equation}\label{eq5}
\Pi_\ell(\lambda)=\frac{1}{N}\int_{-\infty}^{\lambda}\rho_\ell(\lambda')d\lambda',
\end{equation}
for successive generations, as shown in the Figure 1. $\Pi_1$ and
$\Pi_2$ converge very quickly to distinct $g$ independent forms.
Thus, scale free distribution of node degrees and self similarity
do not necessarily lead to NI.

\begin{figure}
\begin{center}
\includegraphics*[width=4.73cm,height=6.6cm,angle=270]{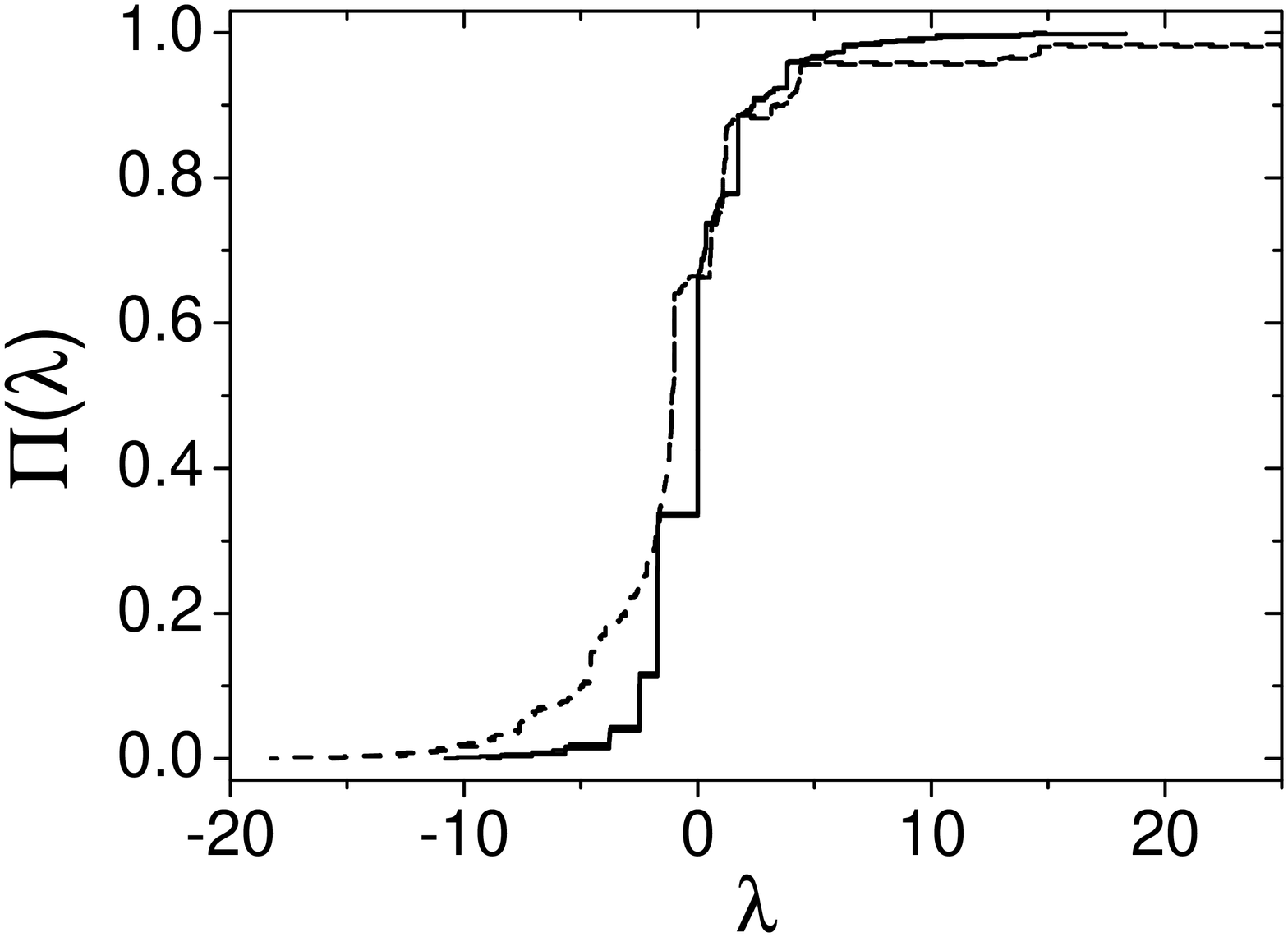}
\end{center}
\caption{Plots of $\Pi\times\lambda$ for $M_1$ (solid) and $M_2$
(dashed) of the AN, when $g=7$ and 8} \label{fig1}
\end{figure}

Exact NI invariance is observed for the linear chain network with
periodic boundary conditions, where each vertex interacts only
with its two next neighbors. The corresponding AM $M_1$ has a well
known pattern, where all 1 elements are placed along the nearest
upper and lower diagonals to the matrix main diagonal. This matrix
has been used to describe a very large number of models in 1
dimension, like the system of phonons, tight-binding electrons
\cite{Ashcroft,Economou}, etc. Also, it has been used as the
starting point to construct small world networks, by changing some
of the original nearest neighbors links according to a given
rewiring probability $p_w$. For $M_1$, $\rho$ is expressed
analytically by the relation
\cite{Ashcroft,Economou,Koiller}
\begin{equation}\label{eq7}
\rho(\lambda)\sim \frac{1}{(4 - \lambda^2)^{1/2}}.
\end{equation}
The other matrices of the family ${M_\ell}$ keep essentially the
same shape, with two sequences of 1's along near-diagonals. Each
one of them moves one step away from the main diagonal upon
increasing the value of $\ell$ by 1, but this does not change the
spectrum. Indeed, this operation can be regarded, e.g. in the
analysis of tight-binding systems, as a decimation procedure of
half of the sites along with a renormalization of the hopping
integral \cite{Koiller}. The resulting system has exactly the same
shape as the original and, consequently, the same spectral
density. The diameter of the network depends linearly on
$\mathbf{N}$, as diagonals move away one step when $\ell$ is
increased by 1.

\begin{figure}
\begin{center}
\includegraphics*[width=4.73cm,height=6.6cm,angle=270]{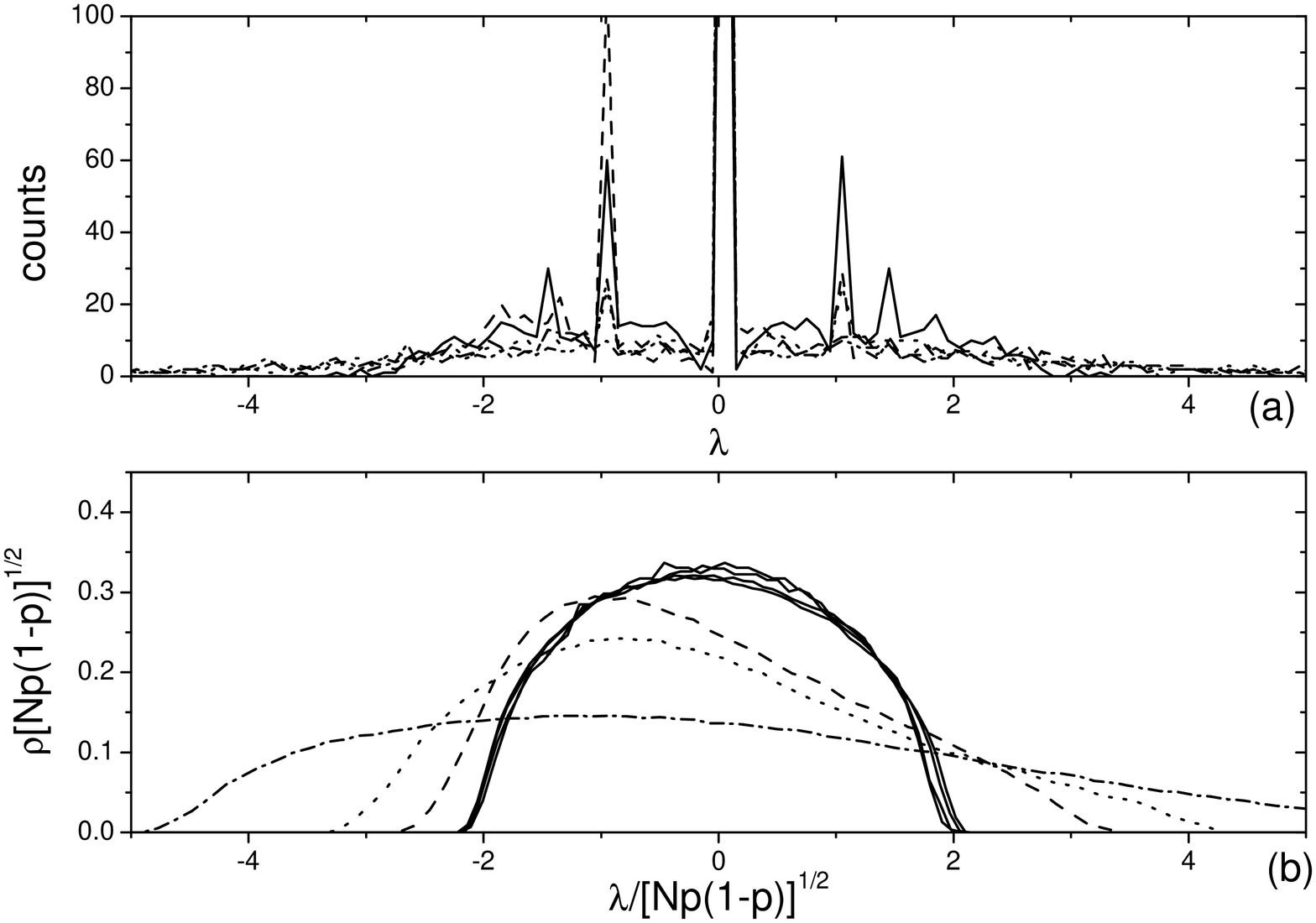}
\end{center}
\caption{Spectral density $\rho_\ell$ for the Erd\"os-R\'enyi
network single realizations. (a) $\mathbf{N}=1000, p=0.05,
\ell=1,2,3,4,5$. (b) Scaled $\rho_2$ for conditions $p=c/N^z$,
when $c=68, z=0.9, N=500, 1000, 3000, 3500$}\label{fig2}
\end{figure}

The Erd\"os-Renyi networks \cite{Renyi} constitute another class
where one could expect to find NI. Indeed, if connections are
randomly distributed for $R_1$, so should they also be for all
members of $\textbf{R}$. For very low values of $p$ (the
connection probability between any two nodes), $R_1$ is split into
several disjoint clusters, and this situation is preserved for all
other $R_\ell$'s. In such cases, the $M_1$ spectral density does
not obey a simple analytical expression, being constituted by some
individual peaks superimposed on a shallow background. The
$M_\ell$ ($\ell\geq 2$) spectra also share the same qualitative
structure. Quantitatively, it is observed an increase of the
$\lambda=0$ dominant peak and a decrease of the other peaks. This
is related to the clustered structure of the network and to the
fact that several clusters are reaching their own diameter.

When $p=c/N^z$, with $z<1$, and for such large enough $c$, almost
all nodes are connected in a single cluster, and $\rho$ obeys the
well known semicircle law \cite{Wigner,Crisanti}
\begin{equation}\label{eq6}
\rho(\lambda)\sim (4 - \lambda^2)^{1/2}.
\end{equation}
In this regime, networks usually have a very small diameter, and
invariance can only be noted for a few values of $\ell$. For the
average node number $\langle k \rangle\simeq 0.5$, we have found
that the $\rho_2$ also obeys (\ref{eq6}), as shown in the Fig. 2b.
However, for smaller values of $\langle k \rangle$, a clear
skewness in the distribution is observed for $\rho_2$, despite the
semicircle form of $\rho_1$. We conclude that Erd\"os-Renyi
networks are NI for a restricted subset of $p$ and $c$.

\begin{figure}
\begin{center}
\includegraphics*[width=4.73cm,height=6.6cm,angle=270]{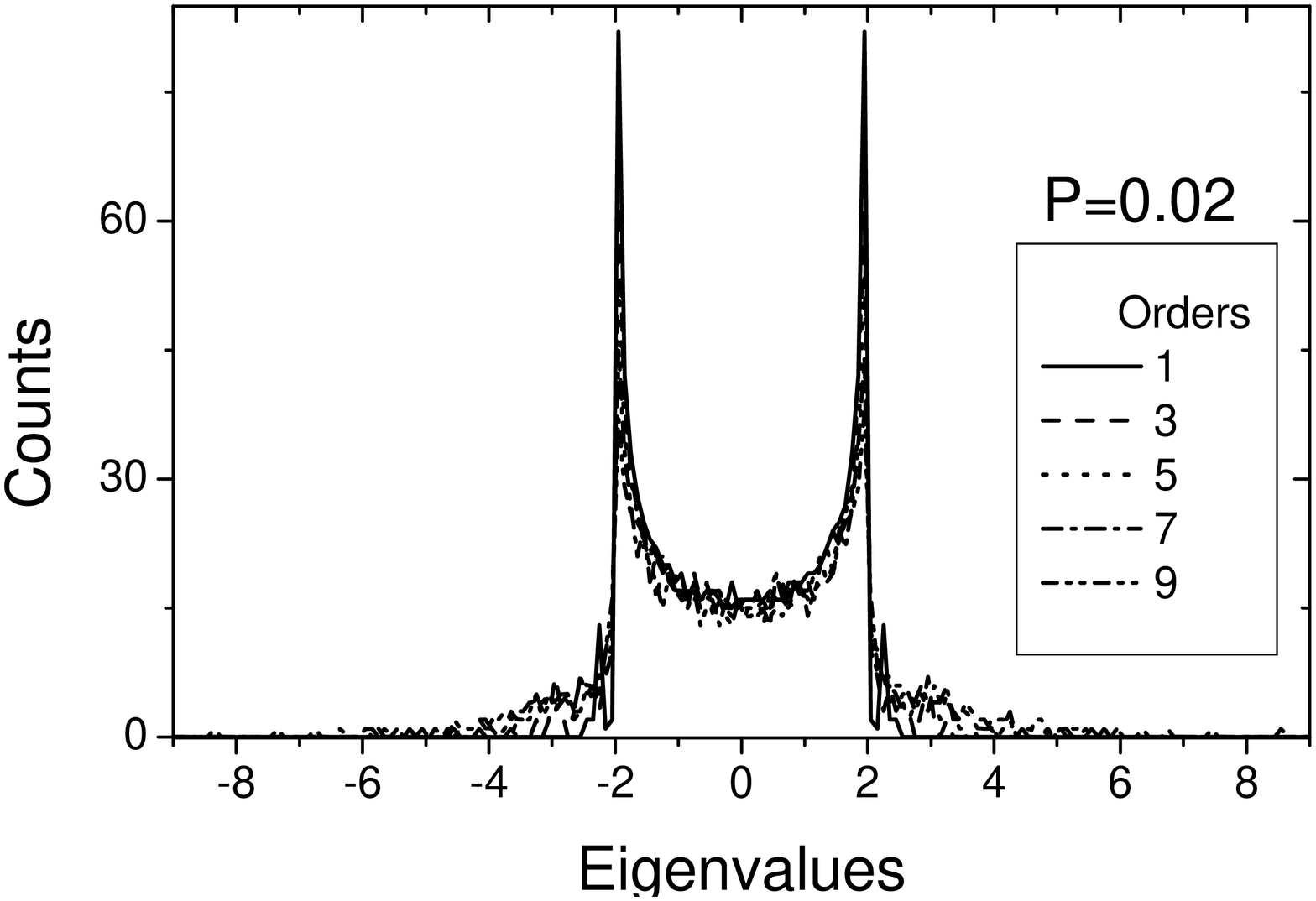}
\end{center}
\caption{Plots of the spectra $\rho\times\lambda$ of the matrices
$M_\ell, \ell=1,3,5,7,9$, for a Watts-Strogatz small world network
starting from a nearest neighbor linear chain and $p_w=0.02$.}
\label{fig3}
\end{figure}
We have also investigated small world networks obtained along the
Watts-Strogatz rewiring procedure\cite{Watts98}, starting both
from the above discussed next-neighbor (nn) and the
nearest-next-neighbor (nnn) linear chains. In Figure 3, we show a
sequence of $M_\ell$ spectra for very small value of $p_w$,
starting from a nn chain. We see that the $\rho(\lambda)$ is split
into two parts: the first one, essentially described by
Eq.(\ref{eq7}), corresponds to the contribution of unperturbed
segments of the linear chain. We see that, as it happens to the
whole spectra when $p_w=0$, this part remains almost invariant, at
least for a large number of $M_\ell$. The other part of the
spectra, located outside of this first region, has no well defined
shape. It entails a considerably lower number of eigenvalues,
which increases with $\ell$. This indicates that, at each increase
of $\ell$, a small number of eigenvalues migrates from the
unperturbed part into it, as the number of vertices that are
affected by the rewiring operation increases with $i$. This
successive migration ends up by affecting the whole form of the
spectrum for large enough $\ell$. Of course this behavior depends
on the value of $p_w$. If we increase it, there is a smooth
transition on the shape of the $M_1$ spectrum, until it reaches a
pattern with many structures and bands, similar to that of the AN.

\begin{figure}
\begin{center}
\includegraphics*[width=4.73cm,height=6.6cm,angle=270]{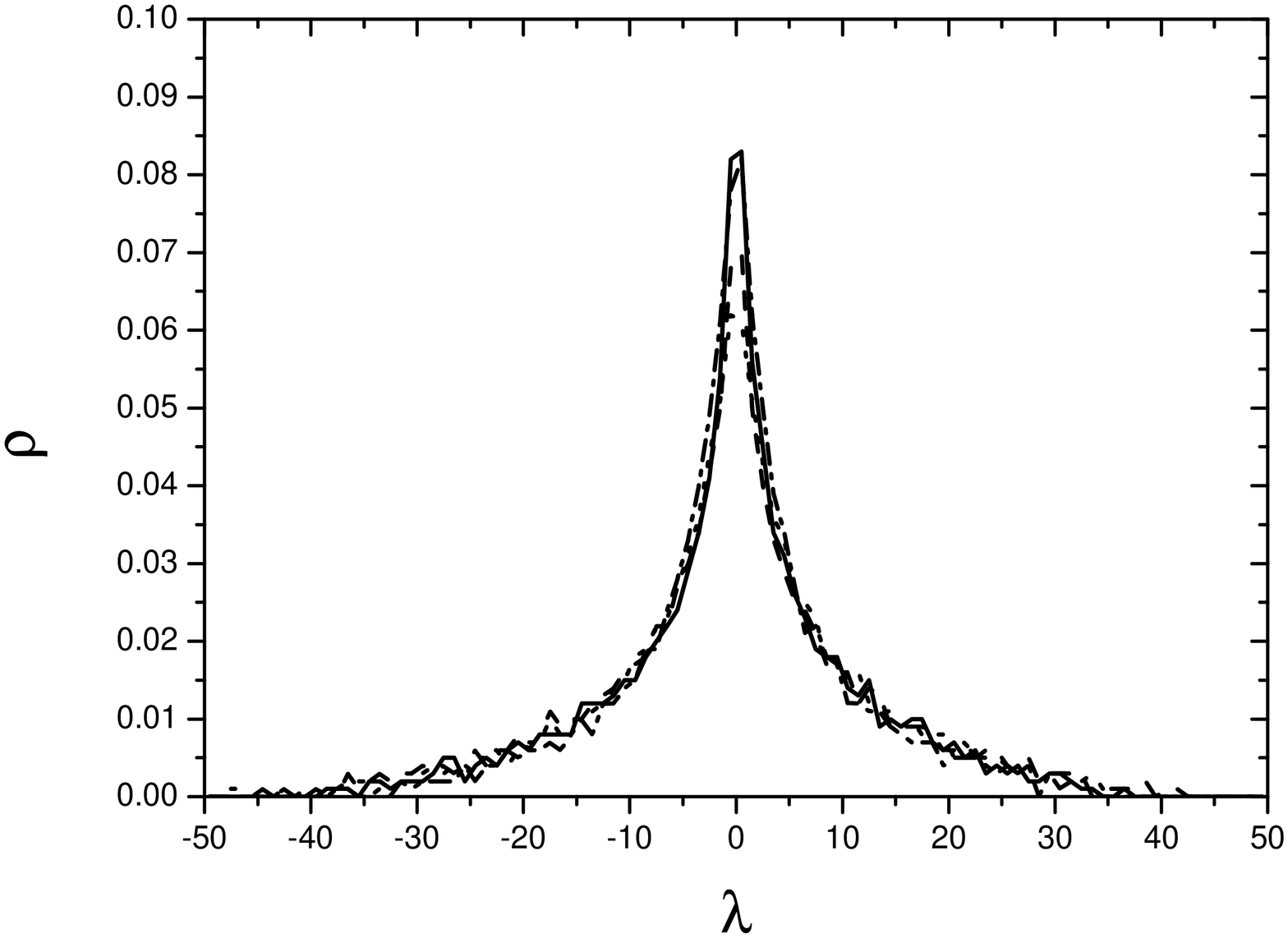}
\end{center}
\caption{Almost invariant spectra $\rho\times\lambda$ of the
matrices $M_\ell, \ell=7-10$, for a Watts-Strogatz- small network
starting from a next-nearest neighbor linear chain and $p_w=0.2$.}
\label{fig4}
\end{figure}

For $p_w$ in the range $p_w\sim 0.15-0.22$, we have observed that,
as $\ell$ increases, the spectral density $\rho_\ell$ evolves
towards a very peculiar form, which remains almost invariant for
several values of $\ell$, as shown in the Figure 4. This spectrum
has its own features, distinct from those discussed before for the
fully ordered and disordered networks.  For the specific case
$p_w=0.21$ and $\mathbf{N}=1000$, we have a large diameter $D=15$,
and the shown form of the spectra remains invariant for $\ell\in
[7,10]$. For smaller values of $\ell$, the shape changes steadily
from a structured shapes similar to those in Figure 3 into the
invariant form. For larger values of $\ell$, finite size effects
lead to quite sparse $M_\ell$, with a large number of zero
eigenvalues: $\rho_\ell$ evolves to a $\delta$ like distribution
centered at $\lambda=0$.

This effect can be graphically illustrated with the help of the
matrix $\widehat{M}$. In Figure 5 we draw the position of the
$O(\ell)$ neighbors for three distinct ranges of $\ell$. The
particular shape in Figure 4 is associated with roughly dense
matrices when $\ell\in [7,10]$. For smaller and larger
values of $\ell$, matrices have rather distinct structure.

\begin{figure}
\begin{center}
\includegraphics*[width=6.75cm,height=9cm,angle=270]{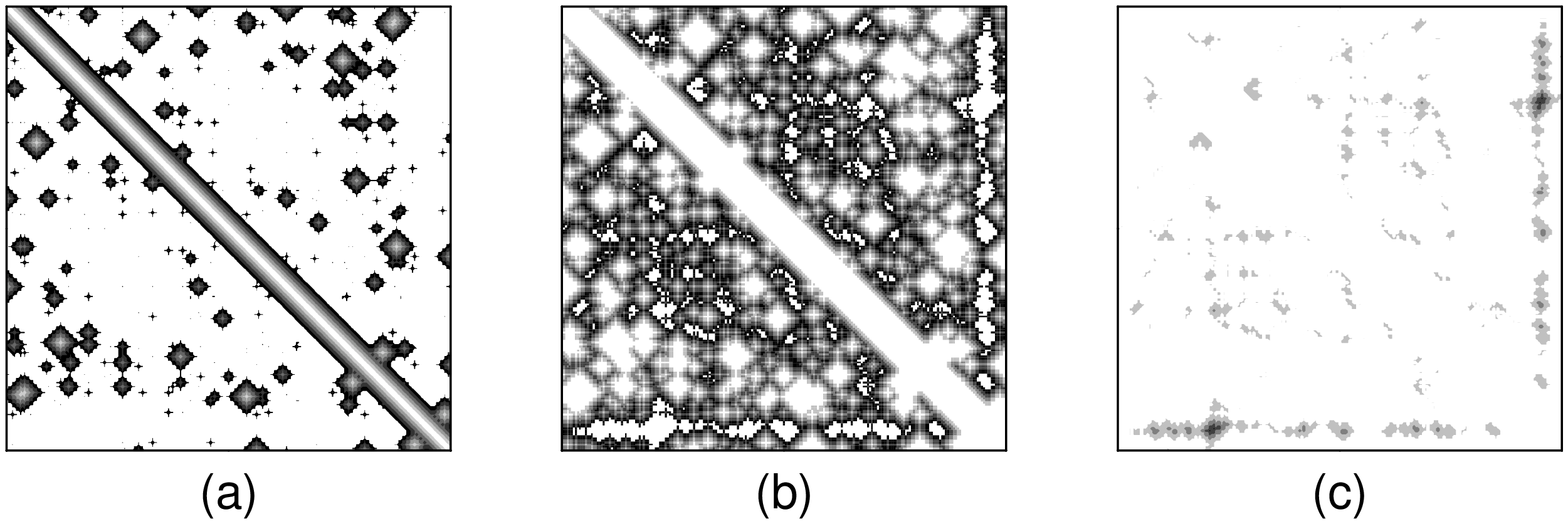}
\end{center}
\caption{For the same Watts-Strogatz small network of Figure 4,
graphical illustration of distribution of $O(\ell)$ neighborhoods:
$\ell\in [1,6], [7,10], [11,15]$ for (a), (b) and (c)
respectively.} \label{fig5}
\end{figure}

For other values of $\mathbf{N}$, we have observed the same
evolution. For instance, when $\mathbf{N}=1500$ and $2000$, the
shape lasts almost invariant for more generations, respectively
$\ell\in [8,12]$ and $[11,16]$, indicating that this behavior can
be more robust as $\mathbf{N}$ increases. For $p_w$ larger than
the range given above, this persistence in the form is not
observed. For smaller values of $p_w$ the spectra changes very
slowly as shown in Figure 3. In such cases, finite size effects
set in prior than any tendency of evolution towards the form shown
in Figure 4.

To conclude, in this work we have discussed the concept of higher
order neighborhood and neighborhood invariance of networks. These
have been obtained by a systematic use of Boolean matrix
operations and the definition of a AM family. We explored
well-know networks, showing that this property is not equivalent
to other concepts of scale and geometrical invariance. Further, we
looked for evidence of NI based on the invariance of the spectral
density, identifying this property in the linear chain,
Erd\"{o}s-Renyi network, and finally, in a non-trivial class that
evolves from Watts small world network.

\textbf{Acknowledgement:} This work was partially supported by
CNPq and FAPESB.

\bibliographystyle{prsty}

\end{document}